\documentstyle[aps,epsf,twocolumn]{revtex}
\input epsf.sty
\begin{document}
\draft
\flushbottom
\twocolumn[
\hsize\textwidth\columnwidth\hsize\csname @twocolumnfalse\endcsname

\title{Effective Field Theory for Layered Quantum Antiferromagnets with
Non-Magnetic Impurities}
\author{Yu-Chang Chen and A.~H.~Castro~Neto}

\address{Department of Physics,
University of California,
Riverside, CA, 92521 }
\date{\today}
\maketitle
\tightenlines
\widetext
\advance\leftskip by 57pt
\advance\rightskip by 57pt

\begin{abstract}
We propose an effective two-dimensional quantum non-linear
sigma model combined with classical percolation theory to study the
magnetic properties of site diluted layered quantum
antiferromagnets like La$_{2}$Cu$_{1-x}$M$_x$O$_{4}$
(M$=$Zn, Mg). We calculate the staggered magnetization at
zero temperature, $M_s(x)$, the magnetic correlation length,
$\xi(x,T)$, the NMR relaxation rate, $1/T_1(x,T)$, and the
N\'eel temperature, $T_N(x)$, in the renormalized classical
regime. Due to quantum fluctuations we find a quantum critical point
(QCP) at $x_c \approx 0.305$ at lower doping
than the two-dimensional percolation threshold $x_p \approx 0.41$.
We compare our results with the available experimental data.
\end{abstract}
\pacs{PACS numbers: 75.10.-b,75.10.Jm,75.10.Nr}

]
\
\narrowtext
\tightenlines

The discovery of high temperature superconductivity in
La$_{2-x}$Sr$_x$CuO$_{4} $ has motivated an enormous number of
experimental and theoretical studies of this and related
materials. La$_2$CuO$_4$ has attracted a lot of 
interest because it is a classical example
of a quantum Heisenberg antiferromagnet (QHAF). La$_{2}$CuO$_{4}$
is a layered quasi-two-dimensional (2D) QHAF, with an 
intraplanar coupling constant $J$ ($J/k_B \approx 1500$ K) 
much larger than the interplanar coupling $J_{\perp }$
($\approx 10^{-5}J$) \cite{chn}. The quantum nonlinear
sigma model (QNL$\sigma$M) is probably the simplest continuum
model with correct symmetry and spin-wave spectrum that reproduces
the low-energy behavior of a QHAF. It has been
successfully used \cite{chn} to explain many magnetic properties of
La$_{2-x}$Sr$_x$CuO$_{4}$ \cite{chubukov}.

In this paper we propose a QNL$\sigma$M allied to classical
percolation theory to study the site dilution effect in
La$_{2}$Cu$_{1-x}$M$_{x}$O$_{4}$, where M is a non-magnetic atom. 
While the theory of disordered classical
magnetic systems is fairly developed \cite{stin} we still lack 
deep understanding of the behavior of the site diluted QHAF \cite{gen}.
As we show below
the interplay between quantum fluctuations and disorder leads to new
effects which cannot be found in classical magnets. In particular
we show that long-range order (LRO) is lost before the system reaches
the classical percolation threshold. Furthermore, we have only two
independent parameters in the theory: the spin-wave velocity
$c_0$ ($\approx 0.74$ eV \r{A}/$\hbar $\cite{lyon}) and the bare
coupling constant $\bar{g}_0$ ($\approx 0.685$ \cite{chn}) of the clean
system ($x=0$). The results for the staggered magnetization,
correlation length, NMR relaxation rate and N\'{e}el temperature
are derived without any further adjustable parameters. 

Our starting point is the 2D site diluted
nearest-neighbor isotropic Heisenberg model
\begin{equation}
H=J\sum\limits_{\left\langle i,j\right\rangle } p({\bf
r}_{i})p({\bf r}_{j}) {\bf S}_{i} \cdot {\bf S}_{j} ,
\label{eq1}
\end{equation}
where $p({\bf r})$ is the distribution function for Cu sites:
$p({\bf r}) = 1$ on Cu sites and $p({\bf r})=0$ on M sites.
Although translational invariance has been lost in (\ref{eq1}),
the Hamiltonian retains the SU(2) invariance for rotations in spin
space. Since the symmetry is continuous Goldstone's
theorem predicts the existence of a gapless mode in the broken
symmetry phase. The ordered phase is characterized by a finite
expectation value of the magnetization,
${\bf n}=\langle {\bf S}({\bf Q}) \rangle$, at the
antiferromagnetic ordering vector ${\bf Q} = (\pi/a_0,\pi/a_0)$
($a_{0}=3.8$ \r{A}).

In the pure system, in accord with the Hohenberg-Mermim-Wagner theorem, 
LRO for a
system with continuous symmetry is only possible at finite
temperatures in dimensions larger than 2.
In the absence of disorder the system has a Goldstone mode which is a spin 
wave around ${\bf Q}$ with energy $E({\bf k})$ and linear dispersion 
relation with
the wave-vector ${\bf k}$: $E({\bf k}) = \hbar c |{\bf k}|$, where $c$ is
the spin-wave velocity. This dispersion relation is a consequence
of the Lorentz invariance of the system. In the paramagnetic phase,
where the continuous symmetry is recovered, all excitations are
gapped because order is only retained in a region of size $\xi$. 
In this case the excitations have dispersion
\begin{eqnarray}
E({\bf k}) = \hbar c \sqrt{{\bf k}^2+1/\xi^{2}} \, .
\label{energy}
\end{eqnarray}

Now consider the case where quenched disorder is present. 
Spin-wave theory, which can only be applied to
(\ref{eq1}) at $T=0$, predicts that Lorentz
invariance is lost even for an infinitesimal amount of impurities
\cite{harris2}. The
dispersion changes to $k \ln(k)$ and 
the spin waves become damped at a rate proportional to
$k$ when $k \to 0$. These results (strictly valid in 2D and $T=0$) are
not directly applicable to the systems in question which 
order at finite temperature \cite{next}. 
At finite temperatures and weak disorder we can consider the criterion
established by Harris for the relevance of disorder in critical
phenomena \cite{hc}. Firstly, we can classify the phase diagram
of the pure system as \cite{chn}: renormalized classical (RC) where
$\xi(T)$ diverges as $\exp(T_0/T)$ (where $T_0$ is a characteristic
temperature scale - see (\ref{xi})); quantum critical (QC) where
$\xi(T) \propto 1/T$; quantum disordered (QD) where $\xi(T) \approx \xi_0$
is constant. If we imagine the pure system being divided into regions
of size $\xi$, each part will have fluctuations in the microscopic
coupling constant ($g$, say) which by the central limit theorem are
proportional to the square root of the number of spins $N(\xi) \propto \xi^2$ 
in that region.
That is, there are statistical fluctuations of order $\delta g(\xi)
\propto 1/\sqrt{N(\xi)} \propto 1/\xi$. On the other hand the thermal
fluctuations in the system are of order 
$\delta T(\xi) \propto 1/\ln(\xi/a_0)$ in RC,
$a_0/\xi$ in the QC and vanishingly small in the QD region. 
For the critical
behavior of the system with weak disorder to be essentially the same as
for the pure system one must require that $\delta T(\xi) \gg \delta g(\xi)$
when $\xi \gg a_0$. Observe that this condition is always fullfilled in
the RC regime and therefore we expect the critical behavior to be the
same as in the pure system, that is, described by a QNL$\sigma$M \cite{chn}. In
the QC and QD regimes the situation is not clear because 
$\delta T(\xi) \sim \delta g(\xi)$ and therefore the effect of disorder
is strong. We conjecture that in these regimes the
critical behavior is different from the one described by a QNL$\sigma$M. 
In this work
we focus entirely in the RC regime.  
Having these results in mind we can
apply classical percolation theory to (\ref{eq1})
\cite{stauffer,harris}. The main parameters of the problem
depend on geometrical factors such as the probability of finding a
spin in the infinite cluster $P_{\infty}(x)$ ($\approx 1-x$, for $x \ll 1$) 
and the bond dilution factor \cite{watson} $A(x)$ ($\approx 1-\pi x+\pi x^2/2$)
(in the expressions below $P_{\infty}(x)$
and $A(x)$ are valid for all $x$ as given by the numerical simulations
\cite{harris}). 
In the classical
case the spin stiffness $\rho_s(x)$ is related to the undoped
stiffness by $\rho_s(x)= A(x) \rho_s(0)$, while the transverse
susceptibility is given by $\chi_{\perp}(x) = (P_{\infty}(x)/A(x))
\chi_{\perp}(0)$ so that \cite{hohenberg} $\rho_s(x) = c^2(x) \chi_{\perp}(x)$.

In this paper we propose an effective field theory which is valid
for $T_N \leq T < J/k_B$ and combines
the Lorentz invariance implied in (\ref{energy}), the Harris criterion
and the results
of percolation theory. In percolation theory, besides
the infinite cluster, we always have finite clusters. A finite
cluster of size $L$ has discrete energy levels and therefore a gap
of order $\hbar c/L$. In what follows we assume $\xi \gg L$ and
ignore the contribution of finite clusters to the magnetic
properties and
focus entirely on the physics of the infinite cluster. It is
obvious from the definition of $p({\bf r})$ that on average
$\langle p({\bf r}) \rangle = P_{\infty}(x)$. Furthermore, site
dilution implies that ${\bf n}^2({\bf r}) = p({\bf r})$. Thus, on
average we have \cite{yana} $\langle {\bf n}^2({\bf r}) \rangle =
P_{\infty}(x)$. In the continuum limit of (\ref{eq1}) the
Harris criterion discussed above indicates that in
the long-wavelength low-energy limit the magnetic properties of
the site diluted problem can be described in terms of an effective 
QNL$\sigma$M:
\begin{eqnarray}
Z=\int D{\bf n} \, \delta\left[{\bf n}^{2}-P_{\infty}(x)\right]
\exp \left\{ -S_{eff}/\hbar \right\}, 
\nonumber
\end{eqnarray}
where
\begin{eqnarray}
S_{eff} &=&
1/2 \int_{0}^{\beta \hbar }d\tau \int
d{\bf r} \left[
\chi_{\perp}(x) \left|\partial_{\tau} {\bf n} \right|^{2}
+ \rho_{s}(x) \left|\nabla {\bf n}\right|^2 \right]
\label{eq3}
\end{eqnarray}
and $\tau$ is the imaginary time direction with $\beta=1/(k_B T)$.
Equation (\ref{eq3}) leads to a natural description of the undoped
system and provides an effective field theory for the QNL$\sigma$M
in the presence of impurities. Moreover, it has incorporated the
correct properties of the classical percolation problem added to
the quantum fluctuations of the QHAF.
It is very simple to show by a change of variables that the action
in (\ref{eq3}) can be rewritten as
\begin{equation}
\frac{S_{eff}}{\hbar} =\frac{1}{2g(x)}\int_{0}^{\beta \hbar c(x)}d\tau
\int d{\bf r}\left( \partial _{\mu }{\bf n}\right) ^{2}
\label{eq4}
\end{equation}
where $g(x) = \hbar c(x)/\rho _{s}(x)$ is the effective coupling
constant of the theory. Moreover, because of the continuum limit
the theory has an intrinsic ultraviolet cut-off $\Lambda(x) 
= 2 \sqrt{\pi P_{\infty }(x)}/a_{0}$ which is fixed by the
total number of states.
In writing (\ref{eq4}) we have not included the topological
term. In a random system one suspects that this term vanishes
as in the pure 2D case \cite{fradkin}. Nevertheless there are
always statistical fluctuations in a random system which are of
order $\sqrt{N_I}$, where $N_I$ is the number of $M$
ions. Thus, the topological term has importance as we discuss
at the end of the paper.

\begin{figure}
\hspace{0cm} 
\epsfxsize=6cm
\epsfysize=6cm
\epsfbox{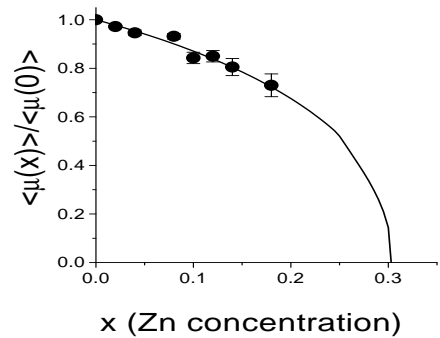}
\caption{Effective moment at $T=0$ as function of x (normalized
relative to the undoped case) and the experimental
data for La$_{2}$Cu$_{1-x}$Zn$_{x}$O$_{4}$
\protect\onlinecite{carretta2}.} 
\label{nqr}
\end{figure}

The great advantage of (\ref{eq4}) is its simplicity and
close relationship to the description of the undoped
problem. In this paper we use the large $N$ approach for the
QNL$\sigma$M which has been so successful in describing the
undoped system \cite{sachdev}. At zero temperature, a critical
value of the coupling constant $g_{c}(x)$ separates the RC 
from the QD region. $g_{c}(x)= 4\pi P_{\infty }(x)/\Lambda(x)$ 
can be obtained from the saddle-point equation for
(\ref{eq4})\cite{sachdev}. The ratio of the coupling
constant to the critical coupling constant is
$\overline{g}(x)\equiv g(x)/g_{c}(x)= \overline{g}_{0}/P_{\infty
}(x)$, which implies that non-magnetic doping drives the system
from RC region to QD region at $x_{c}$ where $P_{\infty
}(x_{c})=\overline{g}_{0}$ at $T=0$. The critical concentration
$x_{c}$ is completely determined by the value of
$\overline{g}_{0}$ in the undoped case. Using the dilute result
for $P_{\infty }(x)$ and $\overline{g}_{0}=0.685$ we find $x_c
\approx 0.3$ which is indeed smaller than the percolation
threshold $x_p \approx 0.41$ \cite{stauffer}. This result has to be
contrasted with classical calculations \cite{rush} where long
range order is lost at percolation threshold only. We also
performed a one-loop renormalization group analysis and calculated
the zero temperature staggered magnetization
\cite{neto} $M_s(x)= M_{0}(x) \sqrt{1-\overline{g}(x)}$. Here
$M_{0}(x)$ is the classical staggered magnetization for the
perfect N\'{e}el spin alignment and the remaining factor is due to
quantum fluctuations. Thus, the local average magnetic
moment is given by
\begin{eqnarray}
\frac{\langle \mu(x) \rangle}{\langle \mu(0) \rangle}
= \frac{M_s(x)/M_0(x)}{M_s(0)/M_0(0)} =
\sqrt{\frac{1-\overline{g}(x)}{1-\overline{g}(0)}} \, .
\label{eq5}
\end{eqnarray}
Observe that the average local moment indeed vanishes at $x_c$.
For the undoped case, (\ref{eq5}) predicts that the maximum
measured magnetic moment of Cu ion is $0.56 \mu _{B}$ which agrees
with the measured value $0.6 \pm 0.15 \mu _{B}$ \cite{vaknin}. It
is also in good agreement with the existing experimental
sublattice magnetization measured by $\mu$SR for various doping
concentrations as shown in Fig. \ref{nqr}. Notice that for the
Ising magnet $\mu(x)$ only deviates from $\mu(0)$ at $x_p$.
The larger reduction of the moment in the QHAF is due to quantum
fluctuations present in the QNL$\sigma$M.

\begin{figure}
\epsfxsize=8cm 
\epsfysize=8cm 
\hspace{0cm} 
\epsfbox{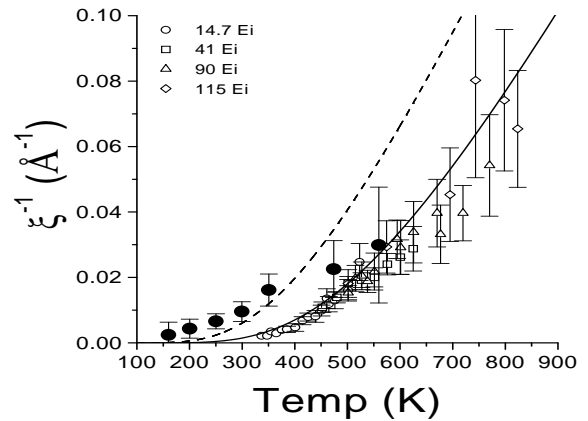}
\caption{Inverse correlation length as function of temperature for
$x=0$ (solid line) and $x=0.05$ (dashed line). The open 
($x=0$) and solid symbols ($x=0.05$) are the neutron scattering
data \protect\onlinecite{keimer}.} \label{xiexp}
\end{figure}

The magnetic correlation length $\xi $ can be directly calculated
from the QNL$\sigma$M. The interpolation formula from the
RC to the QC region reads \cite{neto,hasenfratz}
\begin{equation}
\xi(x,T)=\left( \frac{e\hbar c(x)}{4}\right) \frac{\exp \left(
2\pi \rho_{R,s}(x)/k_{B}T\right) }{4\pi \rho_{R,s}(x)+k_{B}T},
\label{eq6} \label{xi}
\end{equation}
where $\rho_{R,s}(x)=\rho _{s}(x) [1-\overline{g}(x)]$ is the
renormalized spin stiffness. This result
agrees very well with the Monte Carlo simulations in a large
temperature range in the undoped case \cite{beard}.
As far as we know the only existing
neutron scattering results for magnetic correlation length are for
the pure system and $x=0.05$ \cite{keimer}. In Fig. \ref{xiexp} we
plot the available data and the prediction of our model given in
(\ref{xi}). As it is well known, 
samples with $x=0.05$ have problems with the Oxygen stoichiometry \cite{keimer}.
Excess O introduces mobile holes in the plane which produce
strong frustration effects which are not accounted for in our
theory. Thus, direct comparison between the theory and experiment
for this sample is problematic, especially at high temperatures.
Thus, only new experiments with controlled O content can directly test 
our theory.

\begin{figure} 
\epsfxsize=8cm 
\epsfysize=8cm 
\hspace{0cm} 
\epsfbox{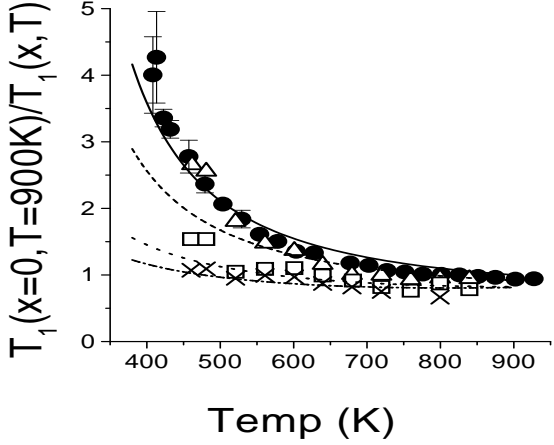}
\caption{The Cu $1/T_{1}$ normalized to
the undoped case at high temperatures ($T=900$ K). 
The lines from top to bottom are for $x=0$
(solid), $x=0.025$ (long dash), $x=0.08$ (short dash), and
$x=0.11$ (dotted dash). $x=0$ (solid circles - 
NQR data \protect\onlinecite{matsumura}) and 
$x=0.025$ (open triangle), $x=0.08$ (open square) $x=0.11$
(cross) (NQR data \protect\onlinecite{carretta}).} \label{nmr}
\end{figure}

Chakravarty and Orbach\cite{orbach} have calculated the nuclear
spin-lattice relaxation rate of Cu for La$_{2}$CuO$_{4}$
using the dynamical structure factor from the QNL$\sigma $M. A
detailed calculation was done in Refs. \cite{carretta,imai}. 
These calculation can be easily extended for the doped case.
Here we just quote the result for $\Lambda \xi \gg 1$:
\begin{eqnarray}
\frac{1}{T_{1}(x,T)}&=&\gamma^{2} P_{\infty}(x)\sqrt{2\pi^{3}}
S(S+1) 
\nonumber
\\
&\times& \epsilon \left(A_{\bot }-4P_{\infty}(x)B\right)
\sqrt{1-\frac{2A_{\bot }B}{A_{\bot }^{2}+4B^{2}}} \nonumber
\\
&\times& \frac{\left[\left(A_{\bot }-4P_{\infty }(x)B\right)
\xi^{2}+4 P_{\infty}(x)B a_{0}^{2} \ln \left( \xi \Lambda
\right)\right] }{3 \omega_e(x) \xi a_{0}\left( \ln \left( \xi
\Lambda \right) \right)^{2}}
\nonumber
\end{eqnarray}
where $\gamma $ is the nuclear gyromagnetic ratio, $A_{\bot }=80$
kG and $B=83$ kG are the hyperfine constants\cite{carretta}, and
\begin{eqnarray}
\omega _{e}(x)=A(x) \sqrt{\left( \frac{2J^{2}k_{B}^{2}zS(S+1)}{
3\hbar^{2}}\right)} 
\nonumber
\end{eqnarray}
(where $z$ is the number of nearest neighbor spins) is the
corrected Heisenberg exchange frequency. Fig. \ref{nmr} shows the
NMR relaxation rate normalized to the high temperature value as
given by the experimental data and the result of our
calculations. The growth of the relaxation rate at low temperatures 
is due to fast growth of $\xi$. 
As the system approaches the QCP one starts to see the crossover from 
RC to the QC regime where
the $\xi$ grows like $1/T$ leading to slower growth of the
relaxation rate. 
This behavior is clearly seen in the data since for
$x=0.11$ where growth is very slow from $800$K down to $400$K. 
The agreement between data and theory is again quite reasonable.

\begin{figure}
\epsfxsize=8cm 
\epsfysize=8cm 
\hspace{0cm} 
\epsfbox{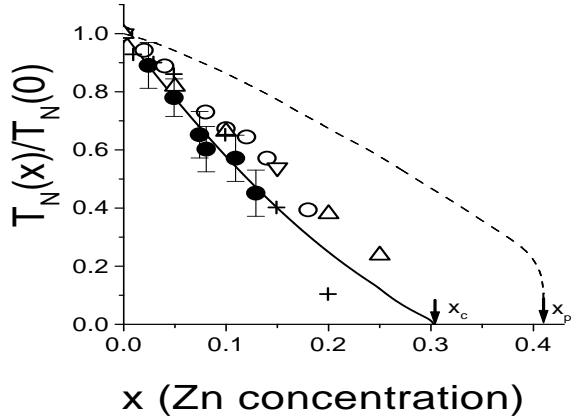}
\caption{ N\'eel temperature normalized to the undoped case. Solid
line: \ref{eq9}; dashed line: Ising result. 
Experimental data for La$_{2}$Cu$_{1-x}$Zn$_{x}$O$_{4}$:
Solid and open circles are NQR and $\mu$SR data,
respectively \protect\onlinecite{carretta}; straight and up-side-down
triangles are
magnetic susceptibility data \protect\onlinecite{hucker}; crosses
are magnetization data \protect\onlinecite{cheong}.} \label{neel}
\end{figure}

The 3D N\'{e}el order can be obtained from the weak
interplane coupling $J_{\bot}$ and it is given by \cite{chn,neto,hone}:
\begin{equation}
k_{B}T_{N} \simeq J_{\bot} P_{\infty}(x)
\left(\frac{\xi(x,T_N)}{a_{0}}\right)^{2}
\left(\frac{M_{s}(x)}{M_{0}(x)}\right)^{2} 
\label{eq9}
\end{equation}
which is a transcendental equation for $T_N(x)$. The interplanar
coupling constant is insensitive \cite{chakraborty} to
doping because the change in lattice parameters is negligible
\cite{uchinokura}. In the undoped case the N\'{e}el temperature $T_{N}(0)$ is
of order of $315$ K. The initial suppression rate of the N\'{e}el
temperature with doping, $I=- d\ln(T_N(x))/dx$, when $x \to 0$ can
be directly obtained from (\ref{eq9}) and, due to quantum fluctuations
it is much faster than
in the Ising case (dashed line on Fig.\ref{neel}). We find $I \approx 4.7$
in good agreement with the data. Indeed, in Fig. \ref{neel} we
show our theoretical results in comparison with various different
experimental measurements. The critical concentration $x_{c}$ for
which the system loses long-range order by moving from the RC
region to the QD region is approximately $0.305$, in agreement
with the loss of long-range order at zero temperature as given in
(\ref{eq5}). Finally, it is also easy  to show using the procedure given
in ref.\cite{nagaosa} that the topological term will lead to
induced moments close to the impurities. These moments interact
through a random magnetic exchange of order $J e^{-(a_0
x)/\xi(x,T)}$. This effect can lead to order of the induced
moments in the paramagnetic phase, 
as seen experimentally \cite{hucker,induce}.

In conclusion, we have proposed an effective QNL$\sigma$M to
describe the magnetic diluted QHAF. Our model combines the result
of classical percolation theory and the quantum fluctuations of
the Heisenberg model. Although our model is fairly simple it gives
a good quantitative description of the magnetism in
La$_{2}$Cu$_{1-x}$M$_x$O$_{4}$. The success of our model in describing
the physics of the RC regime is due to the fact that the 2D correlations
are very long at finite temperatures and the effect of disorder in
the critical behavior is rather weak. Disorder induces quantum fluctuations
in the system which lead to the final destruction of LRO at $x_c$.
This effect is not found in classical magnets where LRO is solely
determined by the percolation problem. Finally, our arguments indicate
that a new approach is required in the QC and QD regions where the
NL$\sigma$M is probably not applicable. 

We thank J.~Baez, W.~Beyermann, F.~Borsa, B.~B\"uchner, P.~Carreta, G.~Castilla,
A.~Chernyshev, M.~Greven, P.~C.~Hammel, B.~Keimer,
D.~MacLaughlin, U.~Mohideen, and S.~Sachdev for useful discussions
and comments. We thank P.~Carretta for providing us with his
experimental results. We also acknowledge support by the
A.~P.~Sloan foundation and support provided by the DOE for
research at Los Alamos National Laboratory.

\begin{references}

\bibitem{chn}S.~Chakravarty {\it et al.},
Phys.~Rev.~Lett. {\bf 60}, 1057 (1988); Phys.~Rev.~B {\bf 39}, 2344 (1989).

\bibitem{chubukov}A.~Chubukov {\it et al.}, Phys.~Rev.~B {\bf 49}, 11919 
(1994).

\bibitem{stin}R.~B.~Stinchcombe in {\it Phase Transitions}, Vol.~7
(Academic Press, London, 1983),pg. 151.

\bibitem{gen}S.~Ting {\it et al.}, Phys.~Rev.~B {\bf 46}, 11772 (1992);
E.~Manousakis, Phys.~Rev.~B {\bf 45}, 7570 (1992).

\bibitem{lyon}K.~B.~Lyon {\it et al.}, Phys.~Rev.~B {\bf 37}, 2353 (1988).

\bibitem{carretta2}P.~Carretta {\it et al.}, Isis Ann.~Rep. A 524 (1996).

\bibitem{harris2}C.~C.~Wan {\it et al.}, Phys.~Rev.~B {\bf 48}, 1036 (1993).

\bibitem{next}Y.~C.~Chen {\it et al.}, unpublished.

\bibitem{hc}A.~B.~Harris, J.~Phys.~C {\bf 7}, 1671 (1974).

\bibitem{stauffer}D.~Stauffer, {\it Introduction to Percolation Theory}
(Taylor \& Francis, London, 1985).

\bibitem{harris}A.~B.~Harris and S.~Kirkpatrick, Phys.~Rev.~B {\bf 16},
542 (1977).

\bibitem{watson}B.~P.~Watson and P.~L.~Leath, Phys.~Rev.~B {\bf 9}, 4893
(1974).

\bibitem{hohenberg}P.~C.~Hohenberg and B.~I.~Haperin, Rev.~Mod.~Phys.
{\bf 49}, 435 (1977).

\bibitem{yana}T.~Yanagisawa, Phys.~Rev.~Lett. {\bf 68}, 1026 (1992).

\bibitem{fradkin}E.~Fradkin and M.~Stone, Phys.~Rev.~B {\bf 38}, 7215 (1988).

\bibitem{keimer}B.~Keimer {\it et al.}, Phys.~Rev.~B {\bf 46},14034 (1992);
R.~J.~Birgeneau {\it et al.}, J.~Phys.~Chem.~Solid {\bf 56}, 1913
(1995); B.~Keimer, Ph.~D. Thesis (MIT, 1991).

\bibitem{matsumura}M.~Matsumura {\it et al.}, J.~Phys.~Soc.~Jpn. {\bf 63},4331
(1994).

\bibitem{carretta}P.~Carretta {\it et al.}, Phys.~Rev.~B {\bf 55}, 3734
(1997);
M.~Corti {\it et al.}, Phys.~Rev.~B {\bf 52}, 4226 (1995).

\bibitem{sachdev}S.~Sachdev, in {\it Low-Dimensional Quantum Field
Theories for Condensed Matter Physicists}, Proc. of the Trieste Summer
School (World Scientific, Singapore, 1992).

\bibitem{rush}G.~S.~Rushbrooke and D.~J.~Morgan, Mol.~Phys. {\bf 4}, 1 (1961).

\bibitem{neto}A.~H.~Castro Neto and D.~Hone, Phys.~Rev.~Lett. {\bf 76},
2165 (1996).

\bibitem{vaknin}D.~Vaknin {\it et al.}, Phys.~Rev.~Lett. {\bf 58}, 2802 (1987).

\bibitem{hasenfratz}P.~Hasenfratz and F.~Niedermayer, Phys.~Lett.~B
{\b 268}, 231 (1991).

\bibitem{beard}B.~B.~Beard {\it et al}., Phys.~Rev.~Lett. {\bf 80}, 1742 (1998).

\bibitem{orbach}S.~Chakravarty and R.~Orbach, Phys.~Rev.~Lett. {\bf 64},
224 (1990).

\bibitem{imai}T.~Imai {\it et al.}, Phys.~Rev.~Lett. {\bf 70}, 1002 (1993).

\bibitem{hone}D.~Hone and A.~H.~Castro Neto, Journal of Superconductivity
{\bf 10}, 349 (1997).

\bibitem{chakraborty}A.~Chakraborty {\it et al.}, Phys.~Rev.~B {\bf 40}, 5296
(1989).

\bibitem{uchinokura}K.~Uchinokura {\it et al.}, Physica B {\bf 205}, 234
(1995).

\bibitem{nagaosa}N.~Nagaosa {\it et al.}, J.~Phys.~Soc.~Jpn. {\bf 65}, 3724
(1996).

\bibitem{hucker}M.~H\"{u}cker {\it et al.}, Phys.~Rev.~B {\bf 59}, R725 (1999).


\bibitem{induce}P.~Mendels {\it et al.}, Phys.~Rev.~B {\bf 49}, 10035 (1994).

\bibitem{cheong}S-W.~Cheong {\it et al.}, Phys.~Rev.~B {\bf 44}, 9739 (1991).
\end{references}
\end{document}